\documentclass[a4paper]{article}

\usepackage{INTERSPEECH2022}
\usepackage{bbm}
\usepackage{xcolor}
\usepackage{multirow}
\usepackage{indentfirst}
\usepackage{subfig}
\title{Adaptive Speech Quality Aware Complex Neural Network for Acoustic Echo Cancellation with Supervised Contrastive Learning}
\name{Bozhong Liu,Xiaoxi Yu, Hantao Huang}

\address{
  MediaTek Singapore}
\email{\{bozhong.liu, xiaoxi.yu\}@mediatek.com, \{huanghantao\}@gmail.com} 

\begin{document}

\newcommand{\comment}[1]{}

\maketitle
\begin{abstract}
Acoustic echo cancellation (AEC) is designed to remove echoes, reverberation, and unwanted added sounds from the microphone signal while maintaining quality of the near-end speaker's speech.  This paper proposes adaptive speech quality complex neural networks to focus on specific tasks for real-time acoustic echo cancellation. In specific, we propose a complex modularize neural network with different stages to focus on feature extraction, acoustic separation, and mask optimization receptively. Furthermore, we adopt the contrastive learning framework and novel speech quality aware loss functions to further improve the performance. The model is trained with 72 hours for pre-training and then 72 hours for fine-tuning. The proposed model outperforms the state-of-the-art performance. 

\end{abstract}

\section{Introduction}

Acoustic echoes can be very annoying since people can hear the echo of own voices. To increase the user experience, the traditional method models the echo impulse repose between the speaker and the microphone as a finite impulse response (FIR) filter, and adaptively adjusts it using normalized least mean square (NLMS) algorithm \cite{benesty2008springer}.  This method works well when it is a far-end single-only scenario. 
For the double talk scenario, where the near-end and far-end speech are present at the same time, the adaptive filter may not converge \cite{dtln_aec}, resulting relatively large volume of residual echoes. 


Deep learning based solution for acoustic echo cancellation has achieved many promising results \cite{fazel2020cad, dtln_aec,zhang2021ft}. For example, \cite{dtln_aec} proposed a DLTN neural network to remove the echo from both the frequency and time domains. However, this architecture lacks the consideration of the phase information. To consider phase information, \cite{zhang2021ft} developed a complex encoder-decoder neural network using complex conv2D and LSTM to build it.  The complex neural network is relatively difficult to compute on hardware (e.g. Neural network accelerator). Instead, a pseudo complex neural network such as DPCRNN \cite{le2021dpcrn}, where it takes the real value and imaginary value as two input channels,  is greatly desired.

Contrastive learning is an effective training method to learn robust representations in the supervised or unsupervised setting \cite{khosla2020supervised,saunshi2019theoretical}. 
The main idea is to use different methods to create positive pairs and negative pairs. The target model is forced to increase and decrease the similarity between positive and negative pairs, respectively. Unsupervised contrastive learning benefits from stronger data augmentation but may lead to worse representation due to false negatives \cite{chen2020simple,khosla2020supervised}. It is due to automatically selected negative pairs that could belong to the same semantic category, creating false negatives. However, existing works on contrastive learning for audio are mainly focused on vector representation. For example, \cite{saeed2020contrastive} proposed a self-supervised pre-training approach
for learning a general-purpose representation of audio for classification based sub-stream tasks such as speaker identification and speech commands. Contrastive learning for regression based tasks has not been fully explored. This paper proposes a new self-supervised contrastive learning for audio regression tasks and uses acoustic echo cancellation as a use-case to demonstrate the performance. 

In this paper, we propose a complex neural network for real-time acoustic echo cancellation. We design a modularize network to perform input feature extraction, speech separation and output predicted speech mask enhancement. Moreover, we propose a supervised contrastive learning loss first time for the regression based tasks.  Furthermore, the novel adaptive speech aware loss functions are proposed to compensate for speech distortion and suppress residual echoes in the speech. Finally, we show the effectiveness of our proposed method through plenty of experiments using ICASSP-2022 AEC dataset \cite{icassp_2022_aec}, which shows that the proposed loss achieves better performance. It improves the PESQ by 0.5 and 0.75 comparing to ICASSP-2022 Microsoft competition baseline \cite{icassp_2022_aec} and DTLN \cite{dtln_aec}.

The broader impact of this work is that the proposed contrastive learning framework can be applied to any regression based audio tasks to improve the robustness. In this paper, we use AEC as an example to show the improvement of robustness. The proposed method can also be used to handle many other problems such as speech enhancement, which will be our future works.

\section{Methodology}
\label{sec:methodology}

\subsection{Problem Formulation}

\begin{figure*}
    \centering
    \includegraphics[scale=0.7]{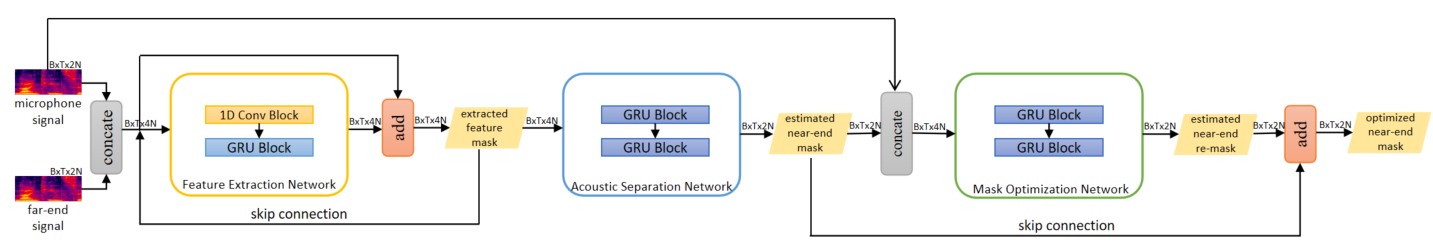}
    \caption{The overall diagram of the proposed network. }
    \label{fig:network structure}
\end{figure*}
We illustrate the signal model of acoustic echo cancellation in the time domain below. The microphone signal $y(n)$ can be formulated as:
\begin{equation}
\label{AEC_eq}
y(n) = s(n) + d(n) + v(n)
\end{equation}

where $s(n)$, $d(n)$, $v(n)$ is  near-end speech, acoustic echo and near-end background noise, respectively. $n$ refers to time sample indexes. $d(n)$ is obtained by the far-end signal $x(n)$ passing through the transmission path $h(n)$. Nonlinear distortion and system delay will be brought in during this process. The acoustic echo cancellation task is to separate $s(n)$ apart from $y(n)$, on the premise that $x(n)$ is known.

\subsection{Architecture}

As illustrated in Figure 1, our deep complex AEC network consists of three modules namely feature extraction network, acoustic separation network and mask optimization network. To use the information of both phase and amplitude of the signal, the input of the network is the stack of the complex form with real and imaginary parts of microphone signal $y(n)$ and far-end signals $x(n)$. 
\subsubsection{Feature Extraction Network (FEN)}
In the feature extraction network, the module is a CGRU structure, which includes a one-dimensional convolution layer (Conv1D) and one gated recurrent unit (GRU) layer. The motivation of such a structure is to extract shallow and deep feature representations of near-end speech and acoustic echo within the sequence of the microphone signal. We build a residual connection by adding the concatenated input with the output of the CGRU and therefore obtain the extracted feature mask. The extracted feature mask is used as the input to the CGRU again. The second-time CGRU output is the complex estimated microphone and far-end signal mask with the same dimension as their input. For one thing, the residual adding and skip connection enhances the input signal and is beneficial to the network allowing the CGRU to better grasp the characteristic information of the signal. For another, passing the repeatable CGRU structure makes better use of the shared weight through the same structure.
\subsubsection{Acoustic Separation  Network (ASN)}
The acoustic separation network is to subtract far-end echo and background noise from the microphone signal. The input is the complex microphone and far-end feature mask from the first network. Two GRU layers are applied to separate the near-end signal information from the input. The output is the estimated near-end speech mask which is half of the input dimension on F (frequency) axis. 
\subsubsection{Mask Optimization  Network (MON)}
In the mask optimization network, both the complex estimated near-end speech mask and original microphone input signal are used as inputs to complex generate a better near-end speech mask, which is added to the coarse complex the input of estimated near-end speech mask to obtain a reﬁned counterpart. For one thing, the estimated mask from the second stage as an input reuses former information. For another, the step-by-step approach can improve performance in complex acoustic scenarios.

\subsection{Contrastive Learning Framework for AEC}
Contrastive learning gains much attention recently in computer vision. It makes the model learn the difference between data pairs to extract better image representation by maximizing the distance of similar data pairs. There are also some models\cite{saeed2021contrastive} that use contrastive learning in a speech to pretrain an encoder to help the down-streaming task obtain a better result. We get inspiration from these algorithms and combine contrastive learning in AEC models to boost the model performance.

\begin{figure}
    \centering
    \includegraphics[width=0.5\textwidth]{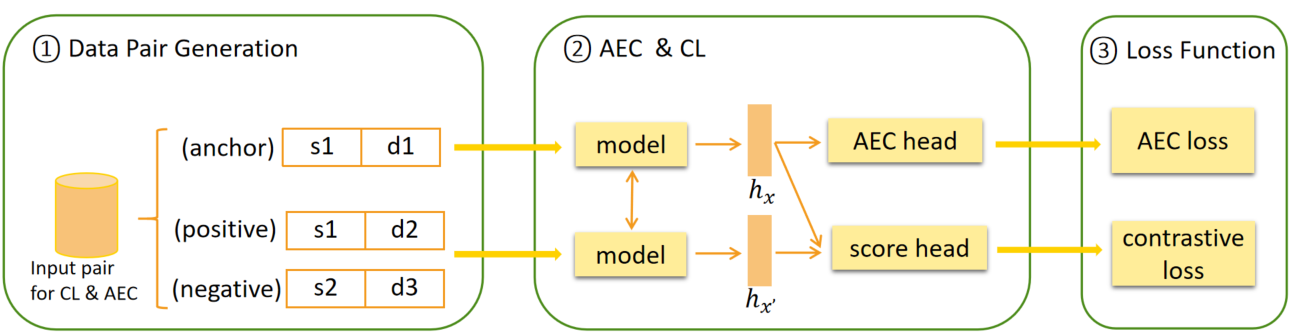}
    \caption{Contrastive learning framework for general ASR (End-to-end speech recognition) }
    \label{fig:framwork}
\end{figure}

{Figure 2} is the framework of our AEC framework with a contrastive learning method. This framework consists of three parts, data pair generation, AEC and CL loss function. 

In the data generation part, we do data augmentation to enable contrastive learning. There is an anchor audio and its corresponding positive and negative audio pair. For positive audio, we use the same near-end audio as anchor audio and different far-end audio. For negative audio, both near-end and far-end audio are different from anchor audio. By maximizing and minimizing the similarity between the positive pair and negative pair, our model can learn to distinguish the near-end audio and far-end audio in the microphone signal.

The augmented data pairs are sent into AEC \& CL part for model training. We use our AEC model with an additional contrastive learning score head to control the training processing of AEC and contrastive learning. The anchor audio and contrastive audio share the same weight as the AEC model. ${h_x}$ and ${h_x'}$ are the features of anchor audio and its contrastive audio pair. The similarity between the anchor feature and contrastive feature is calculated in the scoring head. Only the anchor feature is used in the AEC head. In the loss function part, both AEC loss and contrastive loss are calculated and the total loss of the model is the weighted sum of these two losses as in the equation below.

\begin{equation}
\label{asr_loss}
    \mathcal{L}_{ALL} = \mathcal{L}_{AEC} + \alpha \mathcal{L}_{con}.
\end{equation}
We set $\alpha=1$ in the training stage and $\alpha=0$ in the fine-tuning stage.

While most models calculate contrastive learning loss at the audio level, we compute it at the frame level. The feature of anchor audio and contrastive audio is still in the time-frequency domain after the AEC model. Instead of applying IFFT to convert them back to the time domain, we directly calculate the similarity in frame level. We have m positive audio pair and n negative audio pair for one anchor audio. To calculate the similarity between every pair and anchor audio, it will have (m+n)*T scores and T is the number of frames. In order to reduce the score size, we down-sample the frame size using convolution. Then bilinear comparison\cite{saeed2021contrastive} is performed to get the final similarity score. We use binary cross entropy as a loss function 
\begin{equation}
    \mathcal{L}_{con}=-{\log}\frac{{\exp}(sim(x,x^+))}{\sum_{x^- \in X^-(x)\cup \{x^+\}}{\exp}(sim(x,x^-))}
\end{equation}


\subsection{Loss Function Design}
\subsubsection{Mask Mean Square Error (MaskMSE) Loss}
Both the acoustic separation network and mask optimization network are the process to obtain estimated near-end speech masks, $\hat{M}_{complex}$. Then, the near-end speech can be obtained by multiplying the estimated mask and the microphone signal. We can generate true masks ${M}_{complex}$ by the given true label near-end speech $S(n)$ and microphone signal input $Y(n)$.

The complex mask loss $\mathcal{L}_{M}$ is defined as the mean square error to compare the difference between the estimated complex mask, $\hat{M}_{complex}$ and true mask, ${M}_{complex}$.

\begin{equation}
    {M}_{complex}={M}_{R}+i{M}_{I}
\end{equation}
\begin{equation}
    {M}_R=\frac{{S}_{R}Y_{R}+{S}_{I}Y_{I}}{{Y_{R}}^2+{Y_{I}}^2}
\end{equation}

\begin{equation}
    {M}_I=\frac{{S}_{I}Y_{R}-{S}_{R}Y_{I}}{{Y_{R}}^2+{Y_{I}}^2}
\end{equation}

where {${S}_{R}$, ${S}_{I}$} and {${Y}_{R}$, ${Y}_{I}$} are the real and imaginary parts of s(n), y(n), respectively.
\begin{equation}
\begin{aligned}
    \mathcal{L}_{M}&={|{M}_{complex}-\hat{M}_{complex}|}^2 \\
    &={|{M}_{R}-\hat{M}_{R}+i({M}_{I}-\hat{M}_{I})|}^2
\end{aligned}
\end{equation}
\subsubsection{Error Reduction Loss}
The error reduction loss function ${L}_{error}$ can be introduced to increase the performance of the model. The error signal is defined as the difference between microphone signal input $Y(n)$ and true label near-end speech $S(n)$. Ideally, the same estimated complex mask should reduce all the error signals to zero. Such error is minimized by the multiplication between the modulus of the complex estimated mask and the modulus of the error signal.
\begin{equation}
    {E}_{complex}= {Y}_{complex}-{S}_{complex}
\end{equation}
\begin{equation}
    \mathcal{L}_{error}=|\hat{M}_{complex}||{E}_{complex}|
\end{equation}

\subsubsection{Speech Distortion Compensation Weighted (SDW) Loss}
Speech distortion compensation weight is an adaptive speech quality aware weight to introduce the non-linear weight on MaskMSE. At certain frequency bins, when true mask ${M}_{complex}$ is larger than 1,  it means that clean speech dominates over noisy speech. At such frequency bins, SDW can compensate for speech distortions of estimated speech by applying a weighted power function on MaskMSE.

\begin{equation}
    \mathcal{W}_{SDW}=min({max(1, |{M}_{complex}|)}^n, bound)
\end{equation}

where n and bound are positive numbers \textgreater{1}

\begin{equation}
    \mathcal |{M}_{complex}|=\sqrt{{M}_{R}+{M}_{I}}
\end{equation}

\begin{equation}
    \mathcal {L}_{SDW}={W}_{SDW} \times {L}_{M}
\end{equation}

\subsubsection{Echo Suppression Compensation Weighted (ESW) Loss}
Echo suppression compensation weight is an adaptive speech quality aware weight to introduce the non-linear weight on MaskMSE. At certain frequency bins, when true mask ${M}_{complex}$ is less than 1,  it means that the noisy speech dominates over clean speech. At such frequency bins, the speech signal should be suppressed. ESW can suppress echoes of estimated speech by applying a weighted power function on MaskMSE.

\begin{equation}
    \mathcal{W}_{ESW}=min({min(1, |{M}_{complex}|)}^n, bound)
\end{equation}

where n is a negative number \textless{1} and bound is a positive number

\begin{equation}
    \mathcal {L}_{ESW}={W}_{ESW} \times {L}_{M}
\end{equation}


\begin{table*}[!ht]
\caption{ Performance comparison of different methods at different noise level (SNR) input, in terms of PESQ, ESTOI and ERLE. None stands for the input audio without any processing. Single stands for FEN only. Triple stands for the combination of FEN,ASN, MON. Triple+CL+SQA stands for the triple-contrastive learning framework is trained with adaptive speech quality aware weights including both SDW and ESW. }
\label{results_tbl}
\scalebox{0.7}{
\begin{tabular}{|c|c|ccccc|ccccc|ccccc|}
\hline
\multirow{2}{*}{\textbf{Noise}}                                        & \multirow{2}{*}{\textbf{Model}} & \multicolumn{5}{c|}{\textbf{PESQ}}                                                                                                                                & \multicolumn{5}{c|}{\textbf{ESTOI}}                                                                                                                                & \multicolumn{5}{c|}{\textbf{ERLE}}                                                                                                                               \\ \cline{3-17} 
                                                                       &                                 & \multicolumn{1}{c|}{\textbf{10db}} & \multicolumn{1}{c|}{\textbf{15db}} & \multicolumn{1}{c|}{\textbf{20db}} & \multicolumn{1}{c|}{\textbf{25db}} & \textbf{30db} & \multicolumn{1}{c|}{\textbf{10db}} & \multicolumn{1}{c|}{\textbf{15db}} & \multicolumn{1}{c|}{\textbf{20db}} & \multicolumn{1}{c|}{\textbf{25db}} & \textbf{30db} & \multicolumn{1}{c|}{\textbf{10db}} & \multicolumn{1}{c|}{\textbf{15db}} & \multicolumn{1}{c|}{\textbf{20db}} & \multicolumn{1}{c|}{\textbf{25db}} & \textbf{30db} \\ \hline
\multirow{6}{*}{\begin{tabular}[c]{@{}c@{}}Room \\ Noise\end{tabular}} & None                            & \multicolumn{1}{c|}{2.2652}             & \multicolumn{1}{c|}{2.2873}             & \multicolumn{1}{c|}{2.3100 }             & \multicolumn{1}{c|}{2.3241}             & 2.3346             & \multicolumn{1}{c|}{0.886}             & \multicolumn{1}{c|}{0.887}             & \multicolumn{1}{c|}{0.889}             & \multicolumn{1}{c|}{0.890}             & 0.891             & \multicolumn{1}{c|}{--}             & \multicolumn{1}{c|}{--}             & \multicolumn{1}{c|}{--}             & \multicolumn{1}{c|}{--}             & --             \\ \cline{2-17} 
                                                                       & DTLN\cite{dtln_aec}                            & \multicolumn{1}{c|}{2.766}             & \multicolumn{1}{c|}{2.769}             & \multicolumn{1}{c|}{2.777}             & \multicolumn{1}{c|}{2.777}             & 2.775            & \multicolumn{1}{c|}{0.9019}             & \multicolumn{1}{c|}{0.9019}             & \multicolumn{1}{c|}{0.9024}             & \multicolumn{1}{c|}{0.9027}             & 0.9903             & \multicolumn{1}{c|}{23.73}             & \multicolumn{1}{c|}{23.62}             & \multicolumn{1}{c|}{21.16}             & \multicolumn{1}{c|}{20.80}             & 20.53             \\ \cline{2-17} 
                                                                       & Baseline[10]                        & \multicolumn{1}{c|}{3.0249}             & \multicolumn{1}{c|}{3.0275}             & \multicolumn{1}{c|}{3.0295}             & \multicolumn{1}{c|}{3.0322}             & 3.0308             & \multicolumn{1}{c|}{0.9186}             & \multicolumn{1}{c|}{0.9194}             & \multicolumn{1}{c|}{0.9201}             & \multicolumn{1}{c|}{0.9207}             & 0.9209             & \multicolumn{1}{c|}{22.48}             & \multicolumn{1}{c|}{22.06}             & \multicolumn{1}{c|}{22.03}             & \multicolumn{1}{c|}{22.31}             & 22.30             \\ \cline{2-17} 
                                                                       & Single                          & \multicolumn{1}{c|}{3.0401}             & \multicolumn{1}{c|}{3.0305}             & \multicolumn{1}{c|}{3.0258 }             & \multicolumn{1}{c|}{3.0242 }             & 3.0248              & \multicolumn{1}{c|}{0.9279 }             & \multicolumn{1}{c|}{0.9280 }             & \multicolumn{1}{c|}{0.9287 }             & \multicolumn{1}{c|}{0.9291 }             & 0.9293            & \multicolumn{1}{c|}{24.68}             & \multicolumn{1}{c|}{23.48}             & \multicolumn{1}{c|}{23.06}             & \multicolumn{1}{c|}{22.97}             & 23.04             \\ \cline{2-17} 
                                                                       & Triple                          & \multicolumn{1}{c|}{3.1165}             & \multicolumn{1}{c|}{3.1157}             & \multicolumn{1}{c|}{3.1171}             & \multicolumn{1}{c|}{3.1210}             & 3.1204             & \multicolumn{1}{c|}{0.9307}             & \multicolumn{1}{c|}{0.9310}             & \multicolumn{1}{c|}{0.9317}             & \multicolumn{1}{c|}{0.9320}             & 0.9321             & \multicolumn{1}{c|}{24.68}             & \multicolumn{1}{c|}{23.48}             & \multicolumn{1}{c|}{23.06}             & \multicolumn{1}{c|}{22.97}             & 23.04             \\ \cline{2-17} 
                                                                       & Tripel+CL+SQA                       & \multicolumn{1}{c|}{3.0625}             & \multicolumn{1}{c|}{3.2339}             & \multicolumn{1}{c|}{3.3522}             & \multicolumn{1}{c|}{3.4250}             & 3.4624             & \multicolumn{1}{c|}{0.9302}             & \multicolumn{1}{c|}{0.9378}             & \multicolumn{1}{c|}{0.9424}             & \multicolumn{1}{c|}{0.9451}             & 0.9461             & \multicolumn{1}{c|}{25.24}             & \multicolumn{1}{c|}{23.96}             & \multicolumn{1}{c|}{23.13}             & \multicolumn{1}{c|}{22.76}             & 22.69             \\ \hline
\multirow{6}{*}{\begin{tabular}[c]{@{}c@{}}Music\\ Noise\end{tabular}} & None                            & \multicolumn{1}{c|}{1.5159}             & \multicolumn{1}{c|}{1.5506}             & \multicolumn{1}{c|}{2.3251}             & \multicolumn{1}{c|}{2.3381}             & 2.3473             & \multicolumn{1}{c|}{0.6729}             & \multicolumn{1}{c|}{0.6788}             & \multicolumn{1}{c|}{0.8892}             & \multicolumn{1}{c|}{0.8906}             & 0.8914             & \multicolumn{1}{c|}{1}             & \multicolumn{1}{c|}{1}             & \multicolumn{1}{c|}{1}             & \multicolumn{1}{c|}{1}             & 1             \\ \cline{2-17} 
                                                                       & DTLN\cite{dtln_aec}                           & \multicolumn{1}{c|}{1.8410}             & \multicolumn{1}{c|}{1.8602}             & \multicolumn{1}{c|}{2.7630}             & \multicolumn{1}{c|}{2.7634}             & 2.7630             & \multicolumn{1}{c|}{0.7319}             & \multicolumn{1}{c|}{0.7374}             & \multicolumn{1}{c|}{0.9017}             & \multicolumn{1}{c|}{0.9014}             & 0.9017             & \multicolumn{1}{c|}{1}             & \multicolumn{1}{c|}{1}             & \multicolumn{1}{c|}{1}             & \multicolumn{1}{c|}{1}             & 1             \\ \cline{2-17} 
                                                                       & Baseline[10]                       & \multicolumn{1}{c|}{1.9145}             & \multicolumn{1}{c|}{1.9365}             & \multicolumn{1}{c|}{3.0190}             & \multicolumn{1}{c|}{3.0203}             & 3.0212             & \multicolumn{1}{c|}{0.7499}             & \multicolumn{1}{c|}{0.7570}             & \multicolumn{1}{c|}{0.9191}             & \multicolumn{1}{c|}{0.9196}             & 0.9197             & \multicolumn{1}{c|}{1}             & \multicolumn{1}{c|}{1}             & \multicolumn{1}{c|}{1}             & \multicolumn{1}{c|}{1}             & 1             \\ \cline{2-17} 
                                                                       & Single                          & \multicolumn{1}{c|}{1.9399}             & \multicolumn{1}{c|}{1.9665}             & \multicolumn{1}{c|}{3.0241}             & \multicolumn{1}{c|}{3.0302}             & 3.0332             & \multicolumn{1}{c|}{0.7689}             & \multicolumn{1}{c|}{0.7749}             & \multicolumn{1}{c|}{0.9281}             & \multicolumn{1}{c|}{0.9287}             & 0.9288             & \multicolumn{1}{c|}{1}             & \multicolumn{1}{c|}{1}             & \multicolumn{1}{c|}{1}             & \multicolumn{1}{c|}{1}             & 1             \\ \cline{2-17} 
                                                                       & Triple                          & \multicolumn{1}{c|}{1.9806}             & \multicolumn{1}{c|}{2.0072}             & \multicolumn{1}{c|}{3.0991}             & \multicolumn{1}{c|}{3.1031}             & 3.1019             & \multicolumn{1}{c|}{0.7703}             & \multicolumn{1}{c|}{0.7765}             & \multicolumn{1}{c|}{0.9297}             & \multicolumn{1}{c|}{0.9300}             & 0.9300             & \multicolumn{1}{c|}{1}             & \multicolumn{1}{c|}{1}             & \multicolumn{1}{c|}{1}             & \multicolumn{1}{c|}{1}             & 1             \\ \cline{2-17} 
                                                                       & Tripel+CL+SQA                       & \multicolumn{1}{c|}{3.3743}             & \multicolumn{1}{c|}{3.4631}             & \multicolumn{1}{c|}{3.4983}             & \multicolumn{1}{c|}{3.5099}             & 3.5112             & \multicolumn{1}{c|}{0.8223}             & \multicolumn{1}{c|}{0.8263}             & \multicolumn{1}{c|}{0.8283}             & \multicolumn{1}{c|}{0.8293}             & 0.9399             & \multicolumn{1}{c|}{1}             & \multicolumn{1}{c|}{1}             & \multicolumn{1}{c|}{1}             & \multicolumn{1}{c|}{1}             & 1             \\ \hline
\end{tabular}
}

\end{table*}
\section{Experiments}
\label{sec:experiments}

\subsection{Dataset}

In total, 90h of audio is used, 72h for training, and the remaining 18h for validation. We use clean speech from multilingual data provided by \cite{reddy2021icassp} to generate far-end audio and near-end audio. The dataset contains French, German, Italian, Mandarin, English, Russian and Spanish speech. A detailed description of the original data can be found in \cite{reddy2021icassp}. Before generating far-end and near-end audio, we do speech enhancement using DTLN model \cite{westhausen2020dual} to further clean out the possible noise in the audio.  

For far-end signals and echo signals, random instrumental music noise \cite{snyder2015musan} is firstly added with a signal-to-noise ratio (SNR) taken from a normal distribution with mean 0 and standard deviation 10 to simulate the background noise. To further build a realistic echo signal that reflects the influence of diverse amounts of reverberation, the impulse responses (IR) \cite{ko2017study} are used. The IR dataset contains real impulse responses from various scenarios and simulated impulse responses based on the image methods\cite{allen1979image}. Each IR audio is multiplied by a gain randomly taken from a normal distribution with mean -10 and standard deviation 0 before it is convolved with far-end audio. A high pass filter and band-pass filter are applied to cut off unnecessary frequency in audio. Additional variance and poor transmission characteristics of loudspeakers can be added in this step. The signal is also applied with random spectral shaping with a lower bound of –8/3 and a higher bound of 8/3. To simulate the transmission delay in the device, the signal is delayed by 0-100 ms. We randomly discarded 20\% of the far-end signal and echo signal to create a near-end only scenario in the dataset. 

For near-end signals, we directly use clean speech as near-end audio and convolved it with an echo signal with a signal-to-echo-ratio (SER) taken from a normal distribution with a mean of -10 and standard deviation of 10. For test, SER is randomly chosen in [0,5,10,15].  20\% of near-end audio is also discarded for a near-end-only scenario similar to far-end-only.  

All signals are normalized to (-25,0) dB and segmented to 4 s before passing into the network.

\subsection{Experiment Set-up}
\label{exp setup}
\comment{
To have a fair comparison with our baseline using MAML \cite{winata2020learning}, we adopt the same transformer network with two encoder and four decoder layers. The 6-layer VGG-like CNN module \cite{simonyan2014very} is used to extract information from raw audio. The transformer model is built up with 512 dimension for all layers and has around 10.2M parameters.
We perform a two-stage training strategy for the supervised contrastive learning. The pre-train stage uses both contrastive loss and ASR loss as described in Equation \ref{asr_loss} with learning rate $2.83 \times 10^{-4}$, for $\alpha=1$ and the fine-tune stage uses learning rate $8 \times 10^{-5}$, for $\alpha =0$.  
For the zero-shot setting, the fine-tune stage is performed on the original training dataset (au, en, ir, nz, us) whereas for the full-shot setting, the fine-tune is further performed on the respective dataset (af, hk, in, ph, sg). Note that 100 samples from test dataset are randomly selected for testing, and remaining samples are used for full-shot fine-tune. }


The input of our architecture $y(n)$ is sampled at 16KHz and taken 512 points (32 ms) with 8ms overlapping consecutive frames. After the Fast Fourier transform (FFT), this leads to a 513-dimensional spectral feature in each frame, considering real and imaginary values. The total input features are 1026 including both mic signals and far-end (reference) signals.  The overall neural network architecture constitutes FEN, ASN, and MON.  The network layers are the combination of Conv1D with the filter number 128 and GRU  with the number of units 512.

A two-stage training strategy is performed for supervised contrastive learning. The pre-train stage uses both contrastive loss and MaskMSE loss as described in Equation (3) and  (7) with an initial learning rate of 0.001 while the fine-tuning state uses adaptive speech quality aware (SQW) weighted loss (SDW and ESW) and error reduction loss with initial learning rate 0.00001. The ReduceLROnPlateau learning rate scheduler in Tensorflow is used to monitor the validation loss and if the model was not improved in 5 epochs, the learning rate would be reduced by a factor of 0.1 until it reaches 1e-10. The model is trained with the Adam optimizer \cite{kingma2014adam}.

To evaluate the performance and compare our work with the state of arts, DTLN \cite{dtln_aec} and ICASSP-2022 Microsoft competition baseline \cite{icassp_2022_aec} are our baselines, denoted as baseline in Table 


\section{Results and Discussion}


The performance comparison of different methods at different noise levels (SRN) is summarized in Table in terms of PESQ \cite{pesq}, ESTOI \cite{ETSOI}, and ERLE.  Echo return Loss Enhancement (ERLE) measures the amount of echo loss due to the proposed network, under far-end only scenario.  Perceptual Evaluation of Speech Quality (PESQ) is the evaluation of the speech quality, based on the echo reduced speech after going through the model and the label (ideally clean speech). Extend Short-Time Objective Intelligibility (ESTOI) is another algorithm to evaluate the intelligibility of speech. We have added two kinds of noise: room noise and music noise.  

Our method, single and triple model outperform DTLN \cite{dtln_aec} and ICASSP-2022 Microsoft competition baseline \cite{icassp_2022_aec} under various noise level. For instance, under 30db input noise level, our triple network can achieve 3.204 PESQ values in the room noise compared to DTLN 2.775 and baseline model 3.0275. By further adapting contrastive learning and adaptive speech quality aware (SQA) weight, the performance can be further improved to 3.4624 PESQ values in the room noise and 3.5112 in the music noise under 30db noise of inputs. In addition, the model has been tested on the latest evaluation metric, Perceptual Objective Listening Quality Assessment (POLQA) which provides a new measurement standard for predicting Mean Opinion Scores that outperforms the older PESQ standard. The best model has achieved an average of 3.84 POLQA values under different input noise levels which outperforms the state-of-the-art result (3.22 POLQA).

\begin{figure}
    \centering
        \subfloat[]{\includegraphics[width=0.3\textwidth]{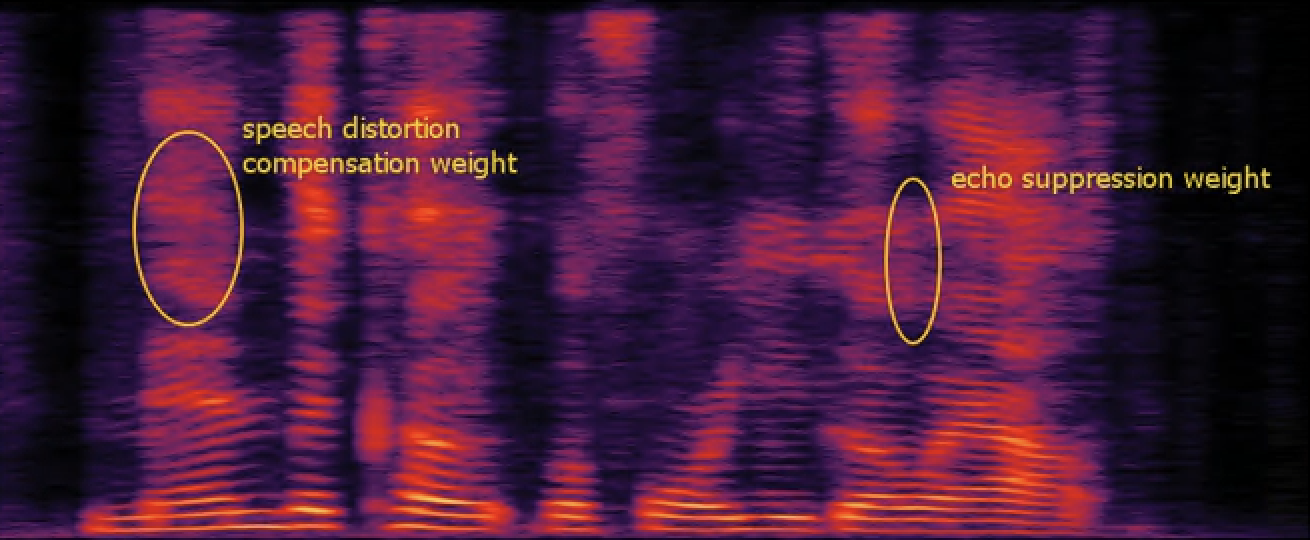}}
    \hfill
        \subfloat[]{\includegraphics[width=0.3\textwidth]{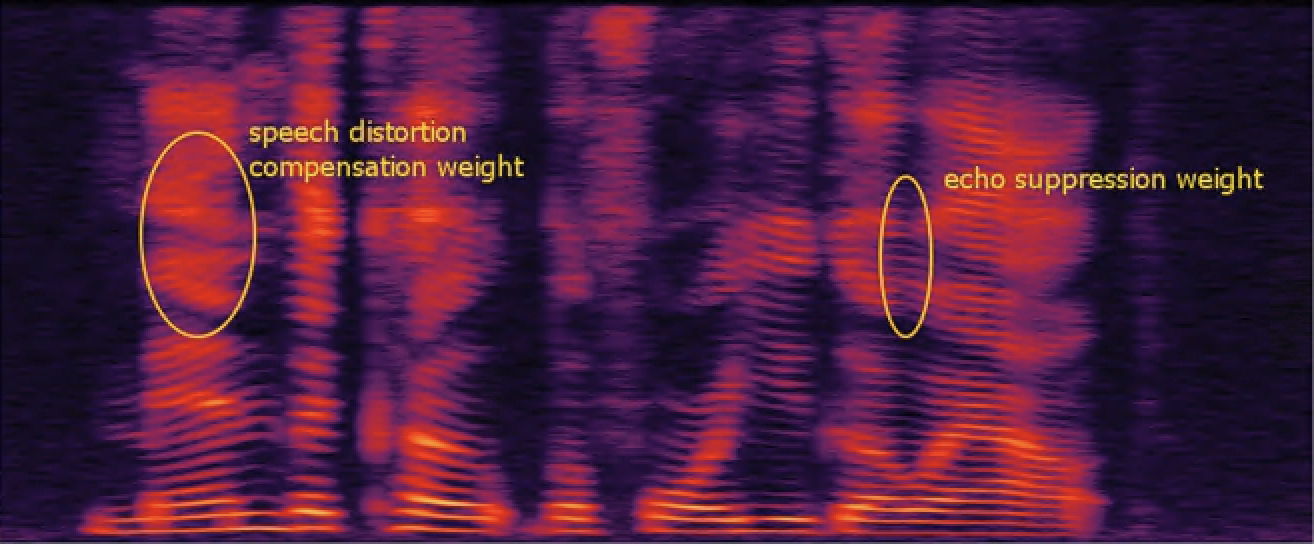}}
\caption{An example of audio spectrum in frequency-time domain before (a) and and after (b) using SQW weights}
\label{fig:before and after spectrum }
\end{figure}

The benefit of adaptive speech quality aware method can be visualized in Figure 3. It compare the difference of an example of enhanced audio spectrum before and after using SQA method. By applying SQA weights, the model can remove the residual echo while maintaining the speech quality by compensate distortion parts  in both low and high frequency band.

\section{Conclusion}
\label{sec:conclusion}
In this paper, we introduced an adaptive speech quality aware complex neural network with a supervised contrastive Learning framework for a robust speech enhancement and echo cancellation system. Three neural network blocks, FEN, ASN, and MON has been proposed and targeted for specific tasks of AEC.  Contrastive learning loss is applied to perform the pre-train and fine-tune without changing model architecture.  The novel loss functions are evaluated under the proposed framework and speech distortion compensation and echo suppression adaptively weighted loss functions benefit the most. Experiments show that contrastive learning with speech quality aware weights can effectively remove speech echoes and improve the speech quality to an average of 3.84 POLQA values under different noise levels which outperforms the state-of-art result.

\vfill\pagebreak
\vfill\pagebreak
\bibliographystyle{IEEEtran}
\bibliography{main}

\end{document}